\documentclass[%
 aip,
 amsmath,amssymb,
preprint,%
]{revtex4-1}

\DeclareSymbolFont{matha}{OML}{txmi}{m}{it}
\DeclareMathSymbol{\varv}{\mathord}{matha}{118}
\usepackage[normalem]{ulem}

\usepackage{graphicx}
\usepackage{dcolumn}
\usepackage{bm}

\usepackage[utf8]{inputenc}
\usepackage[T1]{fontenc}
\usepackage{mathptmx}
\usepackage{etoolbox}

\usepackage[T1]{fontenc}
\usepackage[utf8]{inputenc}
\usepackage[absolute,overlay]{textpos}
\usepackage{hyperref}
\usepackage{graphicx}
\usepackage{dcolumn}
\usepackage{xcolor}
\usepackage{siunitx}
\usepackage[all]{nowidow}

\usepackage{commath} 
\usepackage{braket}  
\usepackage{lipsum} 
\usepackage{import}
\usepackage{color}

\newcommand{\blu}[1]{{\color{black}{#1}}}

\newcommand*{\citen}{}
\DeclareRobustCommand*{\citen}[1]{%
  \begingroup
    \romannumeral-`\x 
    \setcitestyle{numbers}%
    \cite{#1}%
  \endgroup
}

\makeatletter
\def\@email#1#2{%
 \endgroup
 \patchcmd{\titleblock@produce}
  {\frontmatter@RRAPformat}
  {\frontmatter@RRAPformat{\produce@RRAP{*#1\href{mailto:#2}{#2}}}\frontmatter@RRAPformat}
  {}{}
}%
\makeatother

\begin{document}


\title[]{Surfactant-laden Liquid Thread Breakup Driven by Thermal Fluctuations}
\author{Lu\'is H. Carnevale*}
\email{carnevale@ifpan.edu.pl}
\affiliation{Institute of Physics, Polish Academy of Sciences, Al. Lotnik\'ow 32/46, 02-668 Warsaw, Poland}
\author{Piotr Deuar}%
\affiliation{Institute of Physics, Polish Academy of Sciences, Al. Lotnik\'ow 32/46, 02-668 Warsaw, Poland}
\author{Zhizhao Che}
\affiliation{%
State Key Laboratory of Engines, Tianjin University, 300350 Tianjin, China
}%
\author{Panagiotis E. Theodorakis*}%
 \email{panos@ifpan.edu.pl}
\affiliation{Institute of Physics, Polish Academy of Sciences, Al. Lotnik\'ow 32/46, 02-668 Warsaw, Poland}

\date{\today}

\begin{abstract}
The breakup of liquid threads into droplets is crucial in various applications such as 
nanoprinting, nanomanufacturing, and inkjet printing,
where a detailed understanding of the thinning neck
dynamics allows for a precise droplet control. Here, the role of surfactant 
in the breakup process is studied by many-body dissipative particle dynamics, in particular the
various regime transitions and thread profiles, shedding
light on molecular-level intricacies of this process
hitherto inaccessible to continuum theory and experiments. 
Moreover, the role of surfactant in the most unstable
perturbation, the formed droplet size, and surfactant
distributions have been unraveled.
As surfactant concentration rises, both the wavelength and time to breakup 
steadily increase due to the lowering of surface tension below 
the critical micelle concentration (CMC) 
and viscous effects introduced by micelles above the CMC.
These changes prior to the breakup lead 
to larger droplets being formed in cases with higher surfactant concentration.
We also compared the thinning dynamics to existing theoretical predictions, revealing 
that the surfactant-laden breakup starts at the inertial regime and transitions into 
the thermal fluctuation regime when the concentration is increased. 
Thus, we illuminate the hitherto poorly investigated and intricate breakup process of surfactant-laden 
liquid threads driven by thermal fluctuations, contributing to a deeper 
understanding of this process at molecular scales.
\end{abstract}

\maketitle

\section{Introduction}

The breakup of liquid threads into droplets is a ubiquitous natural phenomenon with 
diverse applications,\cite{Moseler2000} such as
nanoprinting,\cite{Basaran2013} nanoscale manufacturing
and chemical processing,\cite{Ye2003} spraying,\cite{wu2021}
and inkjet printing.\cite{Hoath2016} Precise control over droplet size and a comprehensive understanding
of the dynamics surrounding the thinning neck, also known as the thinning bridge, near the
pinch-off point, is crucial in many of these \blu{and other} applications.\blu{\cite{Mousavi2023,Mousavi2022,Mousavi2022b,Heydarpoor2024,Zhao2024}}
The pinch-off process has been extensively examined in the literature, beginning with the
foundational works of Plateau and Rayleigh,\cite{Plateau1857,Rayleigh1878} who delineated the perturbations leading to the
destabilization and subsequent breakup of liquid threads or jets.
Emphasizing surface tension's significance and assuming surface free energy 
minimization, Plateau established the stability condition, i.e. $1 > 2\pi R_0/\lambda = \chi$, where
$\lambda$ represents the perturbation wavelength, $\chi$ signifies the nondimensional 
wavenumber, and $R_0$ denotes the initial radius of the undisturbed fluid. 
Instabilities only arise from perturbations with wavelengths longer than the initial 
circumference of the liquid thread. Rayleigh, via linear stability analysis of inviscid 
liquids, derived the growth rate of such perturbations, identifying the wavenumber 
$\chi= 0.697$ as the point of maximum growth (Rayleigh mode).
Expanding on Rayleigh's work, Weber conducted a similar analysis for viscous fluids, 
establishing that the characteristic wavenumber, determining the highest growth rate, 
relies on the ratio between viscous forces, inertia, and surface tension.\cite{weber1931}
This dependence can be expressed through the non-dimensional Ohnesorge number 
Oh$= \mu/\sqrt{\rho \gamma R_0}$, where $\mu$ is the viscosity, $\rho$ density, and $\gamma$ surface tension.

Given surface tension's pivotal role in liquid thread breakup, it is unsurprising that 
numerous industrial applications utilize surfactants to modify fluid properties and 
enhance process control, particularly in droplet stabilization within emulsions.\cite{goodarzi2019} 
Surfactants, as amphiphilic molecules, have a hydrophobic and a hydrophilic
part. This characteristic leads them to favorably adsorb at the surface of liquids
reducing surface tension. However, there exists a maximum surface concentration, 
$\Gamma_{\infty}$, beyond which 
the fluid surface cannot accommodate more surfactant.
After this point, further increase in the surfactant amount would increase
the number of surfactant monomers in the bulk. As a result, these will come
together and form aggregates of micellar or other morphologies, depending on
the structural characteristics of the surfactant.
The minimum bulk surfactant concentration that
allows for the formation of these aggregates is known as the critical aggregation
concentration (CAC). The more specific term critical micelle concentration (CMC) is 
also widely used, but this term rather indicates that these aggregates are actually micelles.
Importantly, a fluid with surfactants
above the CAC will not experience any more lowering of its surface tension due 
to the saturation of surfactant concentration at the interface. Hence, any further changes in properties above the CAC can be attributed to a higher concentration and the presence of aggregates in the bulk.

During the thinning process of a liquid thread, surfactants are advected away 
from the pinching point and, in this case, there is a competition between the 
advection and the adsorption of surfactants from the bulk toward the interface. 
If adsorption is much slower than advection, a surface concentration gradient is
established that leads to Marangoni stresses that slow down the bridge
thinning.\cite{kovalchuk2018}
Various thinning regimes can occur depending on the balance of forces near the pinch-off.
These regimes are typically characterized by the time before breakup
$\tau = t_b-t$, where $t_b$ is the time that breakup occurs. When inertial forces 
dominate (I regime\cite{eggers1993}), the minimum thread radius at the pinch-off 
region varies as $h_{min}\sim \tau^{2/3}$. 
This result was provided by Eggers, who used self-similar theory and the lubrication 
approximation to simplify the Navier--Stokes equation for an inviscid liquid. He 
has also shown that a universal regime exists when considering 
a balance between viscous and inertial forces (VI regime),
where $h_{min}\sim \tau$.\cite{eggers1993}
There is also an intermediate regime where viscous forces are dominant (V regime),
which also scales as $h_{min}\sim \tau$ as found by Papageorgiou when analyzing 
the pinch-off for the Stokes flow.\cite{papageorgiou1995} 
However, in the V regime only viscous and capillary
forces are relevant with Re depending on time and reaching infinity as pinch-off
is approached. In the VI regime, the inertial forces balance the other forces
with Re number going to unity. For this reason, the V regime may be considered as only an 
intermediate regime, while VI is rather a universal regime from a continuum perspective.
Transitions between these regimes may also take place during the thinning process.
In particular, Castrej\'on-Pita et al. have shown 
the wide range of possible transitions through simulation and experiments.\cite{castrejon-pita2015} 
Moreover, when the adsorption of surfactants is much slower than advection, Wee et al. 
have demonstrated that minimum radius also changes as $h_{min}\sim \tau$, 
following Papageorgiou's solution with 
some corrections to accommodate surface rheological effects.\cite{wee2020} In 
the opposite limit, when adsorption is much faster, the interface remains at 
a constant concentration and, instead of a power-law, the thinning dynamics 
follows an exponential regime where $h_{min}\sim e^{\tau}$. \cite{martinez-calvo2020} 
An exponential regime has also been found in the breakup of viscoelastic fluids.\cite{chang1999}

The different regimes discussed so far are only valid until the minimum radius reaches
a small enough length scale in which thermal fluctuations become relevant, due to the molecular motion. 
This thermocapillary length scale can be defined as $l_T = \sqrt{k_BT/\gamma}$ and depends
on the thermal energy $k_BT$, where $T$ is temperature, $k_B$ Boltzmann's constant,
and $\gamma$ surface tension.
Moseler and Landman\cite{Moseler2000} have studied their influence 
on the breakup process by applying the lubrication approximation to the 
Landau--Lifshitz--Navier--Stokes equation, which contains a stochastic stress tensor
to model the thermal fluctuations and this model is referred to as the stochastic
lubrication equation (SLE). They have found that the surface profile 
near the pinching point becomes a symmetrical double cone which suppresses the formation
of satellite droplets and the model yields predictions comparable to molecular dynamics (MD) simulations.
Eggers has demonstrated that in the thermal-fluctuation thinning regime (TF) 
$h_{min}\sim \tau^{0.418}$ by arguing that surface tension becomes less important 
in driving the breakup and looking for symmetric self-similar solutions to the SLE.\cite{eggers2002} This behavior has been confirmed, by Hennequin et al., from experiments using
ultralow interfacial tension phases of a colloid-polymer solution.\cite{hennequin2006} 
Petit et al. experimentally observed the transition from the VI regime to the TF regime.\cite{petit2012}

Despite the above studies, much less is known about what is
happening when surfactant is present, especially at molecular-level scales.
To investigate the behavior of surfactant-laden liquid threads at this small length 
scale, we use the many-body dissipative particle dynamics method (MDPD).\cite{pagonabarraga2000,pagonabarraga2001,warren2003}
At its core, MDPD employs a coarse-grained representation, grouping particles into
clusters to capture the collective behavior of molecular systems through soft-core
potential interactions, while at the same time, it can be used to simulate
molecular chains, such as surfactants. Combined with 
the ability to model thermal fluctuations, MDPD allows for a computationally 
efficient way to simulate systems at scales that are inaccessible 
to purely macroscopic or atomistic models.
Mostly, MDPD and its predecessor dissipative particle dynamics (DPD) have been successful in describing the breakup
process for surfactant-free systems. For example,
Tiwari et al. \cite{tiwari2008,tiwari2008a} have used DPD and observed 
the symmetric double cone profile shape during pinch-off and were also able to recover
the TF regime minimum radius scaling. Mo et al. \cite{mo2015} have done a more detailed 
analysis of the thinning process using DPD and were able to observe the I regime and 
the TF regime by changing the viscosity and surface tension of the system. They have 
also shown that the viscosity of a DPD fluid is non-Newtonian and this might affect 
the power-law exponent of the V regime. Arienti et al. \cite{arienti2011} have revealed that
MDPD is also capable of describing the breakup process at different coarse-grinning levels and they have also demonstrated the transitions from 
the I to V regime and then to the TF regime. Zhao et al. \cite{zhao2020a} 
used MDPD to simulate fluids with different properties and observed different 
regime transitions when increasing the Oh number and also have proposed 
a new final regime that precedes pinch-off, although no explanation on the nature 
of this regime was given.
Moreover, Zhao et al. \cite{zhao2021} used both MDPD and SLE to study liquid thread 
breakup and obtained the same results, however,
their TF regime scaling differed from Eggers' solution and this discrepancy was attributed
to the neglected influence of surface tension when deriving the power-law.
In our previous work,\cite{carnevale2023} we have investigated different systems 
with MDPD and found that the formation of satellite droplets followed a power-law
that depended on Oh and a thermo-capillary number Th$=l_T/R_0$. 
Despite these studies, we have not been able to find molecular simulations 
of the breakup process
for a surfactant-laden liquid thread. However, different models to simulate systems with
surfactant have been used within the MDPD framework. 
For example, Ghoufi et al.\cite{ghoufi2013}
proposed one of the first models \blu{of} a sodium dodecyl sulfate (SDS)
molecule via a coarse-grained hydrophilic and three hydrophobic particles,
where parametrization of intermolecular interactions is based on the Flory--Huggins theory. 
In another study,
Zhou et al. \cite{zhou2019} modeled SDS and dodecyltrimethylammonium bromide (DTAB) surfactants by first matching the 
surface tension of each particle to experimental values. Then, the intermolecular 
interactions were tuned by constructing different multi-component systems. Recently, 
Hendrikse et al. \cite{hendrikse2023} proposed a different parametrization scheme 
to model alkyl ethoxylate surfactants. In this case, 
they first match the surface tension and density
of each particle to experimental values at a given coarse-grained level and the 
cross-interactions were obtained from the activity coefficients at infinite dilution
computed from the excess chemical potential.

In view of the insufficient fundamental understanding of the breakup process in the 
presence of surfactant at a molecular scale, we have undertaken the task of
exploring this phenomenon in depth
by investigating various key properties.
Specifically, we have carried out MDPD simulations of 
threads that are long enough
to examine how the surfactant concentration changes the characteristic
wavenumber that identifies the perturbation with the highest growth rate leading
to breakup. In addition, we check the difference in sizes of main droplets and 
satellite droplets formed after the breakup for each case considered,
and also
the minimum radius thinning dynamics was investigated and compared with
scaling laws of the aforementioned regimes. Lastly, we present the different 
thread profiles near pinch-off along with the surfactant distribution on the surface
and in the bulk phase, elucidating the surfactant transport mechanism as the system 
evolves towards breakup. Through this thorough investigation, we aim to shed light on the nuanced
molecular mechanisms underpinning this phenomenon, thereby contributing to a more 
comprehensive understanding of surfactant-laden breakup processes at the molecular scale. This
will offer further possibilities in tailored designs for the relevant applications.

\section{Model and Methodology}
\label{model}
\subsection{Many-Body Dissipative Particle Dynamics}

The MDPD model is a mesoscale, particle-based model that evolved from its
predecessor DPD \cite{hoogerbrugge1992,Lavagnini2021} by adding an attractive and a repulsive contribution  
between the particles that depends on their local density. 
This change enables the simulation of systems with 
liquid--vapor coexistence.\cite{pagonabarraga2000,warren2003} MDPD consists in integrating 
the equation of motion Eq.~\ref{eq1} for every particle $i$ that interacts with the
other particles $j$ through a conservative force, $\bm{F}^C$, 
a random force, $\bm{F}^R$, and a dissipative force, $\bm{F}^D$. 
The integration of the equation of motion for each particle $i$ is done by using the 
modified velocity-Verlet algorithm,\cite{groot1997} 
where the equation reads
\begin{eqnarray}
m\frac{d\bm{v}_i}{dt} = \sum_{j\neq i} \bm{F}_{ij}^C + \bm{F}_{ij}^R + \bm{F}_{ij}^D.
\label{eq1}
\end{eqnarray}
The conservative force has a repulsive and an
attractive term that act at different lengths. Its most common form is
\begin{eqnarray}
\bm{F}^C_{ij} =  A\omega^C(r_{ij})\bm{e}_{ij} + 
B \left(\bar{\rho_i} + \bar{\rho_j} \right) \omega^d(r_{ij})\bm{e}_{ij},
\label{eq2}
\end{eqnarray}
where $A<0$ and $B>0$ are the attractive and repulsive parameters, respectively, 
$r_{ij}$ is the distance between particles, $\bm{e}_{ij}$ is the direction vector
from particle \textit{i} to particle \textit{j}, $\omega^C(r_{ij})$ and $\omega^d(r_{ij})$ are linear weight functions defined as follows
\begin{eqnarray}
\omega^{C}(r_{ij}) = 
\begin{cases}
&1 - \frac{r_{ij}}{r_{c}}, \ \ r_{ij} \leq r_{c} \\
& 0,  \  \ r_{ij} > r_{c},
\end{cases} 
\label{eq3}
\end{eqnarray}
\begin{eqnarray}
\omega^{d}(r_{ij}) = 
\begin{cases}
&1 - \frac{r_{ij}}{r_{d}}, \ \ r_{ij} \leq r_{d} \\
& 0,  \  \ r_{ij} > r_{d},
\end{cases} 
\label{eq3.2}
\end{eqnarray}
with $r_c$ being a cutoff distance for the attractive interaction, 
usually set to unity, while the repulsive interaction cutoff is usually $r_d=0.75r_c$.
Although not necessary for our purposes, changing $r_d$ affects 
the parameter space $(A,B)$ where liquid, vapor, and solid phases are present.\cite{Vanya2018}

The many-body contributions in the repulsive force that come from the 
dependence on local densities $\bar{\rho_i}$ and $\bar{\rho_j}$ are calculated as
\begin{eqnarray}
\bar{\rho_i} = \sum_{j\neq i} \frac{15}{2\pi r_d^3} \left( 1 - \frac{r_{ij}}{r_d}\right)^2 .
\label{eq4}
\end{eqnarray}
Other functions can be used to compute the local densities, such as the kernel
functions commonly used in the smoothed particle hydrodynamics (SPH) method.\cite{wang2016}

To incorporate thermal fluctuations in MDPD, random and dissipative forces are introduced and act as a thermostat, keeping the temperature in the simulation constant and equal to unity.
Both can be expressed as
\begin{eqnarray}
\bm{F}^D_{ij} = -\sigma \omega^D(r_{ij}) (\bm{e}_{ij} \cdot  \bm{v}_{ij})\bm{e}_{ij} ,
\label{eq5}
\end{eqnarray}
\begin{eqnarray}
\bm{F}^R_{ij} = \xi \omega^R(r_{ij}) \theta_{ij} \blu{\Delta t^{-1/2}} \bm{e}_{ij} ,
\label{eq26}
\end{eqnarray}
where $\sigma$ is the dissipative strength, $\xi$ is the strength of the random force, 
$\bm{v}_{ij}$ is the relative velocity between particles, $\theta_{ij}$ is a random 
variable from a Gaussian distribution with \blu{zero mean and} unit variance. \blu{
$\Delta t$ is the timestep, taken to be equal to $0.01$}.
According to the fluctuation--dissipation theorem,\cite{espanol1995} $\sigma$ and 
$\xi$ are related to each other by
\begin{eqnarray}
\sigma = \frac{\xi ^2}{2 k_B T}, 
\label{eq7}
\end{eqnarray}
and the weight functions for the forces are
\begin{eqnarray}
\omega^D(r_{ij}) = \left[\omega^R(r_{ij})\right]^2 = \left( 1 - \frac{r_{ij}}{r_c}\right)^2.
\label{eq8}
\end{eqnarray}
\noindent
For multi-component systems, $A=A_{ij}$ is the attraction parameter between 
particles of type $i$ and of type $j$. The repulsive parameter $B$ has
to be the same for all interactions due to the no-go theorem 
otherwise the force wouldn't be conservative.\cite{warren2013} Other types of 
local density functions might be used to circumvent this restriction, 
however, they have to be carefully defined to avoid unphysical behavior 
in the simulations.\cite{vanya2020}

\subsection{Parametrization}

\begin{figure}[bt!]
\centering
\includegraphics[width=0.45\textwidth]{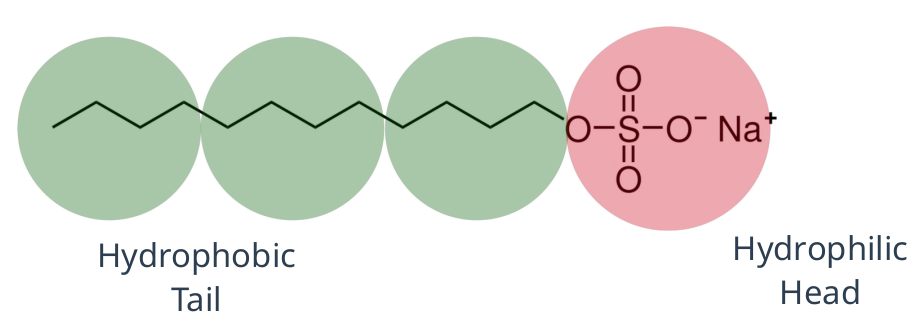}
\caption{\label{fig:sds} Schematic representation of the coarse-grained $HT_3$ surfactant
model (see text for details).\cite{zhou2019} }
\end{figure}

To simulate a physical system with different components in MDPD, it is necessary 
to define the coarse-graining level, i.e., how many atoms or molecules are represented
by one MDPD particle. A typical choice is to define one particle as three 
water molecules and match the model reduced units to real units by measuring 
some key properties such as density and surface tension for a given set of 
simulation parameters. In this paper, we have adopted the commonly used values
$A_{ww} = -40$ and $B=25$ (MDPD units), 
where the subscript `\textit{w}' denotes the self interaction
between water beads\blu{, and the dissipative coefficient $\sigma=4.5$}.\cite{Ghoufi2011, arienti2011} The conversion between reduced and real units is detailed in Table \ref{tab:units}.
With the interaction between our liquid particles being set, we follow the parametrization
of sodium dodecyl sulfate (SDS) surfactant molecules in the same manner as 
Zhou et al.,\cite{zhou2019} which was chosen due to its ample use in industrial 
processes. At the coarse-grained level, the molecule is represented by 
one `H'-type particle, which is the hydrophilic head group of the surfactant 
and by three `T'-type particles, which model the alkane hydrophobic tail of the molecule.
Figure \ref{fig:sds} shows, schematically, the coarse-grained SDS model.

\begin{table}[b]
\caption{\label{tab:units} Conversion between MDPD units and real units. The scaling 
is done by matching surface tension and density of water to values measured from MDPD
simulations using $A=-40$ and $B=25$. The coarse-graining level is defined so that
one MDPD particle represents three water molecules.}
\begin{ruledtabular}
\begin{tabular}{lll}
Parameter  & MDPD value  & Real value \vspace{.1cm} \\ 
 \hline 
Particle &  1     & 3 H$_2$O  \\ 
$r_c$    &  1     &  8.17 \AA          \\ 
$\rho$   &  6.05  &  997 kg/m$^3$       \\
$\gamma$ &  7.62  &  72 mN/m     \\

\end{tabular}
\end{ruledtabular}
\end{table}

The self interaction between T particles $A_{TT}$ is tuned to reproduce the 
surface tension of hexadecane and the cross interaction between $w$ and T particles
$A_{wT}$ was adjusted with the interfacial tension between water and hexadecane (T-type particles). 
The self and cross interactions with the H-type particle were fitted by also 
measuring the interfacial tension between hexadecane and water with SDS molecules at
the interface. All interactions are summarized in Table \ref{tab:values}.
Moreover, we used harmonic potentials to model the bonds and angles
between particles in the surfactant molecule in order to preserve the structure
of the molecule. 
We found that the bond length and strength proposed by Ref.~\citen{zhou2019} were not 
stable in our simulations, 
so we adopted the values $r_0 = 0.45$, $k=400$ of Ref.~\citen{zhu2021}, 
while the same angle potential parameters\cite{zhou2019} were kept, namely, $\theta_0=160^\circ$, $k=100$.

\begin{table}[b]
\caption{\label{tab:values} MDPD interaction parameters for the HT$_3$ surfactant model
and water. W is the water particle, H the hydrophilic particle, and T the hydrophobic particle. The MDPD repulsive parameter is the same for all interactions and is $B=25$.}
\begin{ruledtabular}
\begin{tabular}{cccc}
 $A_{ij}$  & W  & H  & T \vspace{.1cm} \\ 
 \hline \vspace{.3cm}
W   &  -40     &        &       \\ \vspace{.3cm}
H   &  -32.18  &  -19   &       \\
T   &  -27     &  -5.98 &  -22  \\

\end{tabular}
\end{ruledtabular}
\end{table}

\section{Results and Discussion}
\label{results}

\subsection{Surfactant properties}

\begin{figure}[bt!]
\centering
\includegraphics[width=0.5\textwidth]{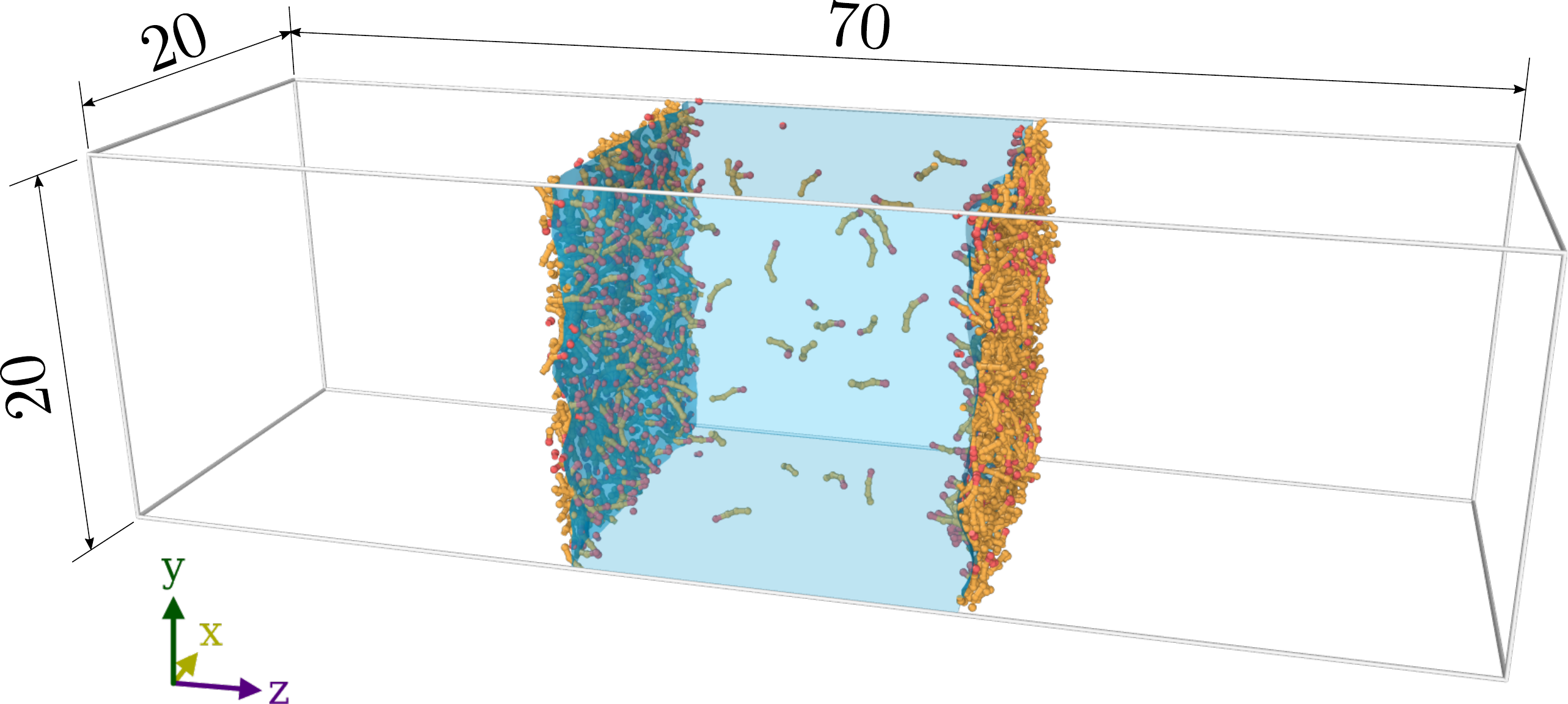}
\caption{\label{fig:st-sim} Example of a simulation box used to measure the surface tension 
of a liquid slab in the presence of surfactants. 
The liquid particles are represented 
by the shaded blue volume for easier visualization.
In this case, the concentration of surfactants is below CMC and the surfactant molecules
are not aggregated in the bulk phase.}
\end{figure}

We have conducted a series of simulations to comprehensively characterize the pertinent 
attributes of our system when surfactants are introduced. The primary focal point of our validation
investigation pertains to the surface tension, specifically its change by varying
surfactant concentration. Figure~\ref{fig:st-sim} shows the standard
setup used in simulation to measure the surface tension based on the 
Kirkwood--Buff method.\cite{Kirkwood1949}
Since the CMC is reached when the interface is saturated,  the initial 
surfactant concentration can be defined as the total number of surfactant molecules $N_t$ divided 
by the initial surface area of the system $A_s$, namely, $C = N_t/A_s$, while the surface excess 
concentration is obtained by counting only the number of molecules on the surface divided 
by the surface area, $\Gamma = N_s/A_s$.
Lastly, bulk concentrations are also presented in
terms of the surface area $C_b = N_b/A_s$, and, therefore, $ N_t = N_b + N_s$.
\begin{figure}[htpb]
\centering
\includegraphics[width=0.5\textwidth]{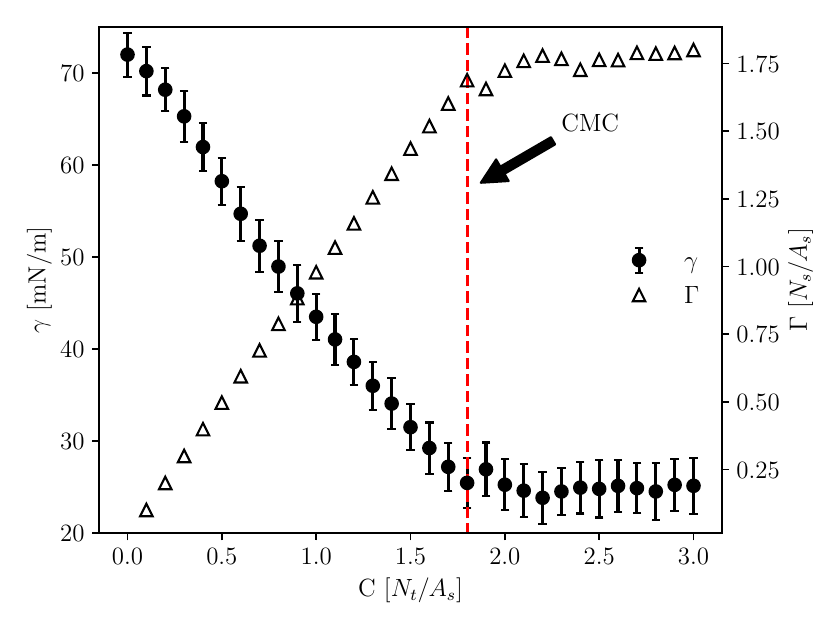}
\caption{\label{fig:st-result} Dependence of surface tension, $\gamma$ (circles) 
and surface excess concentration, $\Gamma$ (triangles), 
on the number of surfactant molecules per surface area, $C$. 
Both values reach a plateau after crossing the CMC (dashed line) indicating that the
interface is fully saturated.
}
\end{figure}
Our simulations for determining the surface tension for the water--SDS-surfactant system start
with the initialization of a water slab with a thickness $h=20$ \blu{and particle number density
$\rho=6.05$} at the midpoint of the periodic simulation box. 
This box was defined by dimensions of $L_x=L_y=20$ and $L_z=70$ (Figure~\ref{fig:st-sim}).
Subsequently, surfactant molecules were positioned directly at the interfaces 
of this water slab, and we allowed the system to run $10^5$ time steps to attain
thermodynamic equilibrium. Once equilibrium was reached,
we performed an additional $10^6$ time steps of simulation to acquire data 
for calculating the properties of the system.
A representative configuration of an equilibrated system is visually depicted in Figure~\ref{fig:st-sim}. To determine the surface tension, we employed the Kirkwood--Buff method,\cite{Kirkwood1949} i.e.,
\begin{equation}
    \gamma = \frac{L_z}{2}\left(P_{zz} - \frac{P_{xx}+P_{yy}}{2} \right)
    \label{eq:surface-tension}
\end{equation}
where $P_{zz}$ is the normal component of the pressure tensor, while $P_{xx}$ and
$P_{yy}$ are the components of the tangential directions to the liquid--vapor surfaces.
An equally significant parameter is the surface-excess concentration of surfactants. 
These compounds have a natural propensity to adsorb at the liquid interface 
until reaching a point of saturation, 
where the surface-excess concentration attains its maximum, while surface
tension $\gamma$ reaches its minimum.
Beyond this saturation threshold, the introduction of additional surfactants
leads to their aggregation within the bulk liquid, forming the micelles.
In Figure \ref{fig:st-result}, we present the findings that illustrate the variation 
of both the surface tension and the surface excess concentration 
in response to differing surfactant concentrations. 
In addition, the CMC point is clearly discernible in our results and shown in the plot. 

\begin{figure}[bt!]
\includegraphics[width=0.5\textwidth]{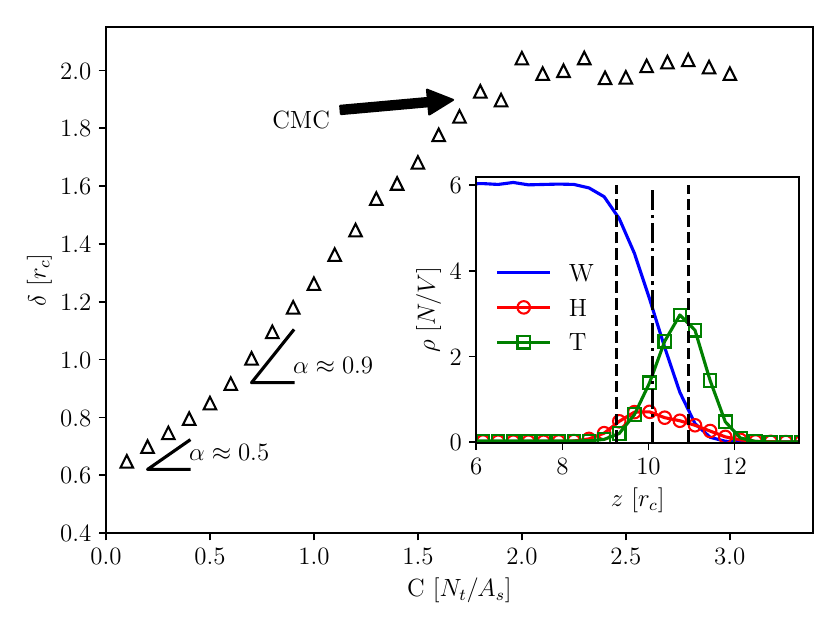}
\caption{\label{fig:st-density} Change in the interfacial thickness with the increase in 
the number of surfactant molecules per surface area ($\alpha$ gives the slope.). It remains at a constant value upon 
reaching the CMC. Inset: example of a density distribution near the interface for a simulation 
below CMC. The vertical lines indicate the position of the interface and its thickness from 
a hyperbolic tangent fit (Eq.~\ref{eq:hyperbolic_tangent}). W, H, T are the liquid,
head, and tail beads, respectively. 
}
\end{figure}

The interfacial thickness is another important quantity in describing the region 
affected by surfactants along a fluid's interface.
This measurement can be obtained by fitting the density distribution of
$w$ particles across the interface using the hyperbolic tangent function
\begin{equation}
\label{eq:hyperbolic_tangent}
    \rho (z) = \frac{\rho_l}{2}\left[ 1-\tanh{\left( \frac{2(r-R_0)}{\delta}\right)} \right]
\end{equation}
where $\rho_l$ represents the bulk density of the liquid phase, $R_0$ is the Gibbs 
dividing surface position and $\delta$ the thickness of the interface. 
An example of such a distribution 
is depicted in the inset plot of Fig.~\ref{fig:st-density} for a simulation with
$C=1.5$. We can see how the density of $w$ particles changes from the liquid phase
where $\rho = 6.05$ to the vapor phase with $\rho \approx 0$ and the vertical lines indicate
the Gibbs dividing surface (dash-dot) and the thickness of the interface (dashed). 
The density of each surfactant particle is also notable and it shows that the head group 
H tends to stay closer to the liquid phase while the tail particles are much closer to 
the vapor phase.
The primary plot in Fig.\ref{fig:st-density} showcases the variation in interfacial thickness.
As surfactant concentration rises, $\delta$ increases until it reaches
a plateau value when $C=1.8$, indicating the CMC saturation point. 
For concentrations below $0.6$, the thickness increases linearly with slope 
$\alpha \approx 0.5$ and between $C=0.6$ and the CMC with slope $\alpha \approx 0.9$.
These two linear regimes can be explained by the orientation of the surfactant molecules
with respect to the interface. At lower concentrations, the surfactant tails stay 
in contact with the liquid surface, while at a higher concentration they become more 
packed orienting in the normal direction to the surface
due to the decrease in available surface area per molecule.\cite{hendrikse2023}
In this sense, the cross-over at $C=0.6$ reflects the range of the interactions
between individual surfactant molecules at the fluid surface.

\subsection{Wavenumber and droplet sizes}

\begin{figure}[bt!]
\centering
\includegraphics[width=0.5\textwidth]{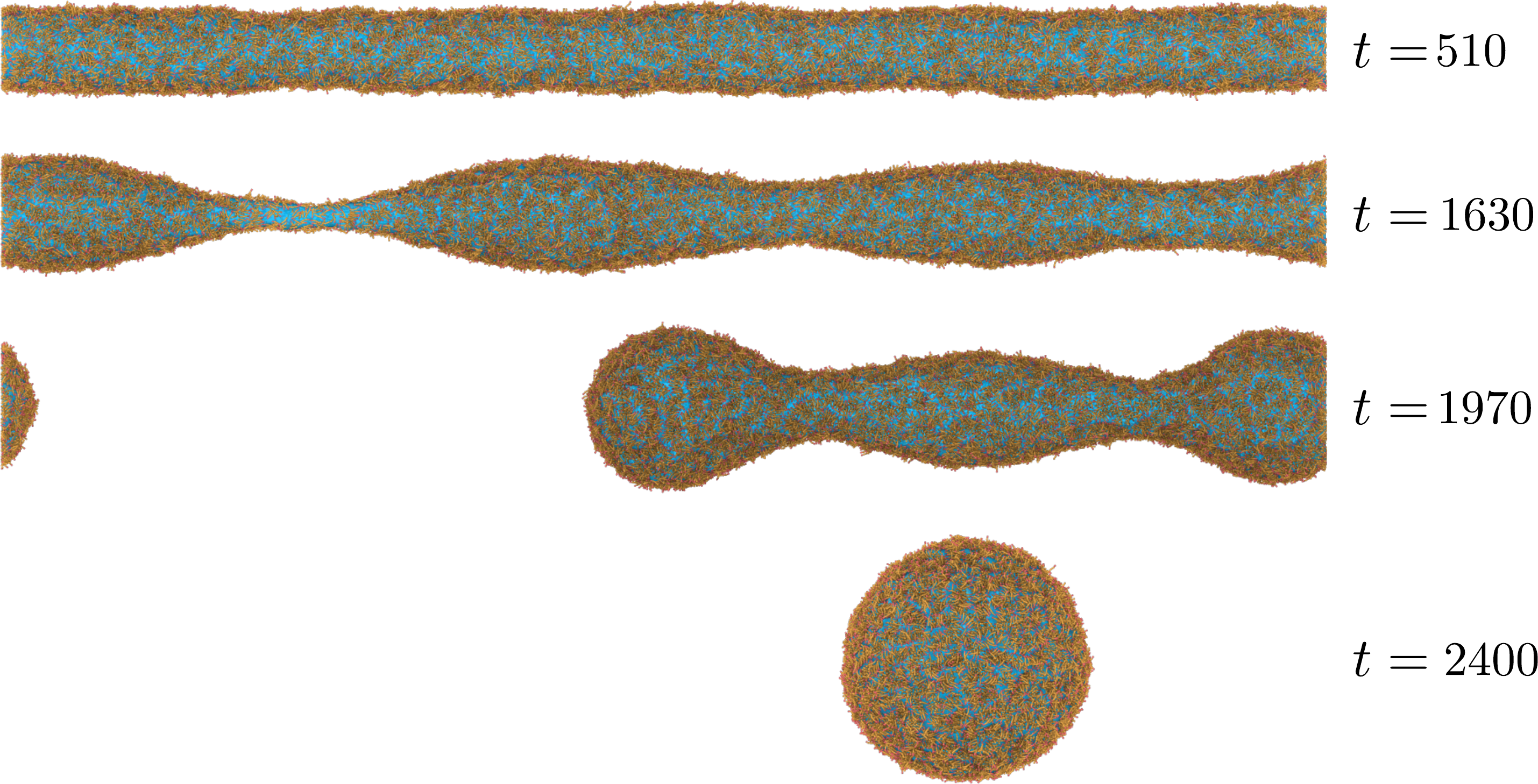}
\caption{\label{fig:breakup-sim} Simulation snapshots of the breakup process
of a small liquid thread. 
The system develops perturbations that grow with time and lead to the pinch-off 
and subsequent formation of a droplet. The concentration of surfactants is below CMC in
this example. The threads used for statistical analysis were four times longer in the 
axial direction than the simulation shown here, which was shortened for the sake of clarity.}
\end{figure}

Following our previous work with pure liquid threads,\cite{carnevale2023}
we have determined the characteristic wavelength of the
breakup process, expressed by the reduced wavenumber $\chi =2\pi R_0/\lambda $. 
This is done by computing the density correlation function along the
$z$ direction,\cite{Theodorakis_2009} that is the direction along the thread.
To construct our initial configuration, we first make a water cylinder 
with unperturbed radius $R_0=8$ and length $L_z=24$, which is 
stable ($L_z < 2\pi R_0$) and does not break. 
In turn, we add the surfactant molecules on its surface, according to the 
desired concentration and we run the system for $2\times10^5$ time steps to reach equilibrium. 
After reaching equilibrium, we replicate the system in the 
$z$ direction $48$ times to obtain a long thread and let it evolve 
in time until it breaks up and all droplets are formed. 
We have already verified that possible finite-size effects in the
direction along the thread quickly disappear, but long threads enable a
more accurate calculation of the characteristic length scale of the
breakup process when realizing the Fourier transform of the density--density
correlation, as has already been shown for pure liquid droplets.\cite{carnevale2023}
Moreover, long threads naturally lead to the formation of a
larger number of droplets, which in turn allows for better statistics on the properties of
the main and satellite droplets. We realized 20 such simulations 
for each surfactant concentration and the averaged results are discussed in this study.
A time sequence of snapshots that 
leads to the formation of a single droplet is presented in Figure~\ref{fig:breakup-sim}. 
In this example, a shorter thread simulation is being shown just for clarity as 
the longer threads would be poorly represented on a plot due to its dimensions.
To form this thread, we replicate the equilibrated system $12$ times in the axial direction, giving thread length $L=12L_z$.
It is clearly visible that a perturbation starts to develop with $\lambda \approx 
L/3 
$, but, due to asynchronous breakup, only one final droplet is formed,
a mechanism that has been explained in detail in our previous work with pure liquids.\cite{carnevale2023}  

\begin{figure}[bt!]
\centering
\includegraphics[width=0.5\textwidth]{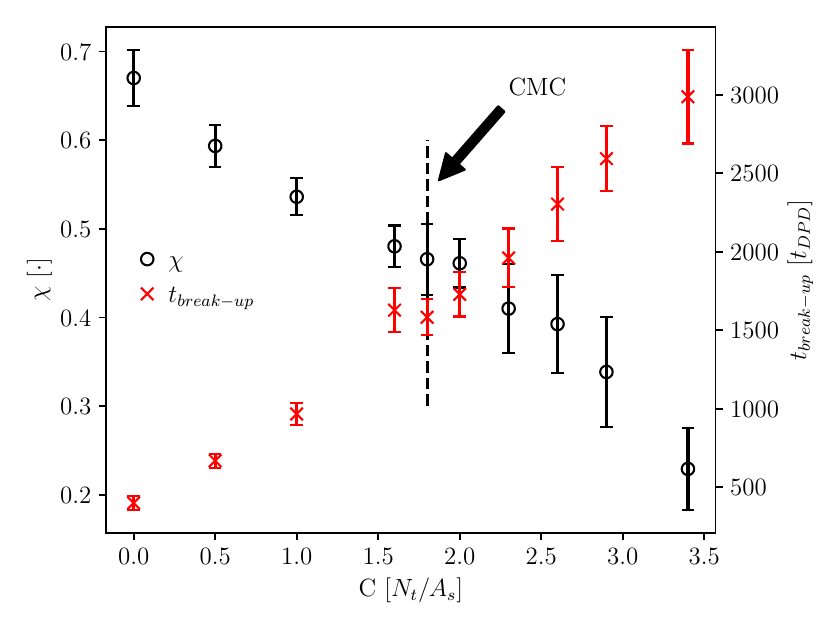}
\caption{\label{fig:wavenumber} Change in reduced characteristic wavenumber and time to breakup vs total 
surfactant concentration (total number of surfactant molecules divided by 
the initial surface area of the system). }
\end{figure}

As indicated by our results, increasing surfactant concentration leads to a notable 
reduction in the surface tension. Since surface tension acts as the primary 
force of the instability, its decrease subsequently decelerates the overall 
breakup process. This is visually evident in Figure \ref{fig:wavenumber}, 
showcasing a discernible growth in the time required for the thread to break ($t_{break-up}$),
which shows a linear growth with surfactant concentration, also beyond the CMC.
Moreover, the characteristic wavenumber, $\chi$ (see Ref.~\citen{carnevale2023} for details on calculating $\chi$),  decreases as more surfactant is
added to the system, which means that perturbations with longer wavelengths become more 
effective in destabilizing the liquid thread. However, this decrease is 
not only due to the change in surface tension caused by the surfactants 
as $\gamma$ remains constant for concentrations above the CMC. 
Thus, the formation of micelles also adds viscous effects
in the liquid bulk which explains the further slow-down of the breakup process above CMC.
We also observe an increase in the standard deviation in our measurements, 
which can be attributed to the increase of the thermocapillary length 
consequently intensifying the significance of fluctuations within the system.

\begin{figure}[bt!]
\centering
\includegraphics[width=0.5\textwidth]{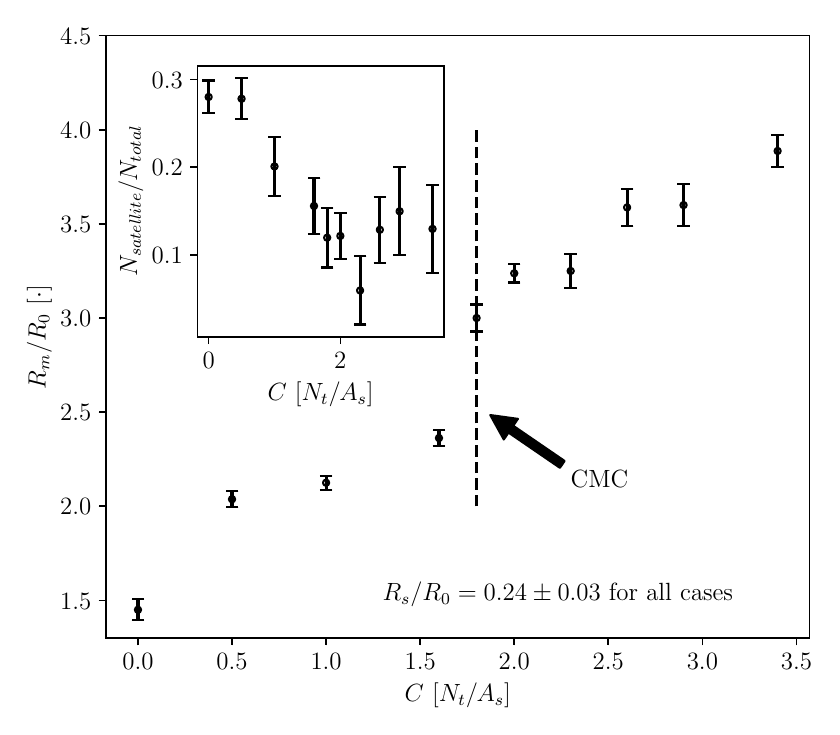}
\caption{\label{fig:dropsize} Size of main droplets $R_m$ normalized with the
initial thread radius vs total surfactant concentration, $C$. 
The radius of satellite droplets did not present any statistically significant variation with respect 
to the change in concentration, and therefore
data are now plotted here (see main text for further details). 
Inset shows the change in the proportion between the number of 
satellite droplets and the total number of droplets formed.}
\end{figure}

After the breakup, we track the number of main and
satellite droplets over time by doing a cluster analysis with the software OVITO.\cite{ovito}
The clusters were defined as a group of particles that are close to each other
within a threshold value and the droplet sizes were obtained from the radius of 
gyration of spherical clusters following the protocols 
of our previous work.\cite{carnevale2023}
Figure \ref{fig:dropsize} shows the increase in size of the main droplets 
formed as surfactant concentration increases. 
This change in size is expected as the wavelengths that lead to the breakup also increase, 
subsequently forming fewer pinching points along the thread axis. There is a slight 
difference between systems below and above CMC
in how the main droplet sizes increase. 
This can be explained by the slowing down of the breakup 
process, which favors the suppression of pinching in a few regions of the thread, 
similar to what is depicted in Figure \ref{fig:breakup-sim}.
Furthermore, we could not observe any statistically significant variation in the satellite droplet size with concentration, it
having values in the range $R_s/R_0=0.24\pm 0.03$ in all the simulated cases.
However, there is a clear reduction in the proportion in which they are formed
as $C$ increases
that could also be attributed to the larger thermocapillary length scale, since 
thermal fluctuations are believed to inhibit their formation.\cite{Moseler2000}
This proportion approximately attains an asymptotic  
value above CAC as shown in the 
inset of Figure~\ref{fig:dropsize}.

\subsection{Surfactant distribution at the liquid--vapor interface}

Since the breakup is influenced by surface tension, describing how the surfactant
is distributed along the interface is of great importance in understanding the 
pinch-off process and the formation of droplets.
To capture these distributions, we investigated two systems below and two above the CMC, conducting 20 simulations for each system. These simulations were done with short liquid threads where $L_z = \lambda_c = 2\pi R_0/\chi$, ensuring breakup driven by the most unstable perturbation.
Utilizing OVITO software,\cite{ovito} we separated the particles into axial bins and tracked
interface particles. Circles were fitted to these particles' positions in each bin, with their radii defining the interface profile $h$. Additionally, surfactant molecules within each bin were quantified, and assigned to the surface if within an interface thickness $\delta$ from $h$ or otherwise to the bulk phase. The number of molecules on the 
surface $N_s$ and in the bulk $N_b$ are divided by the surface area $A$ on each 
bin to obtain the surface excess $\Gamma$ and the bulk concentration $C_b$.

\begin{figure}[bt!]
\centering
\includegraphics[width=0.70
\textwidth]{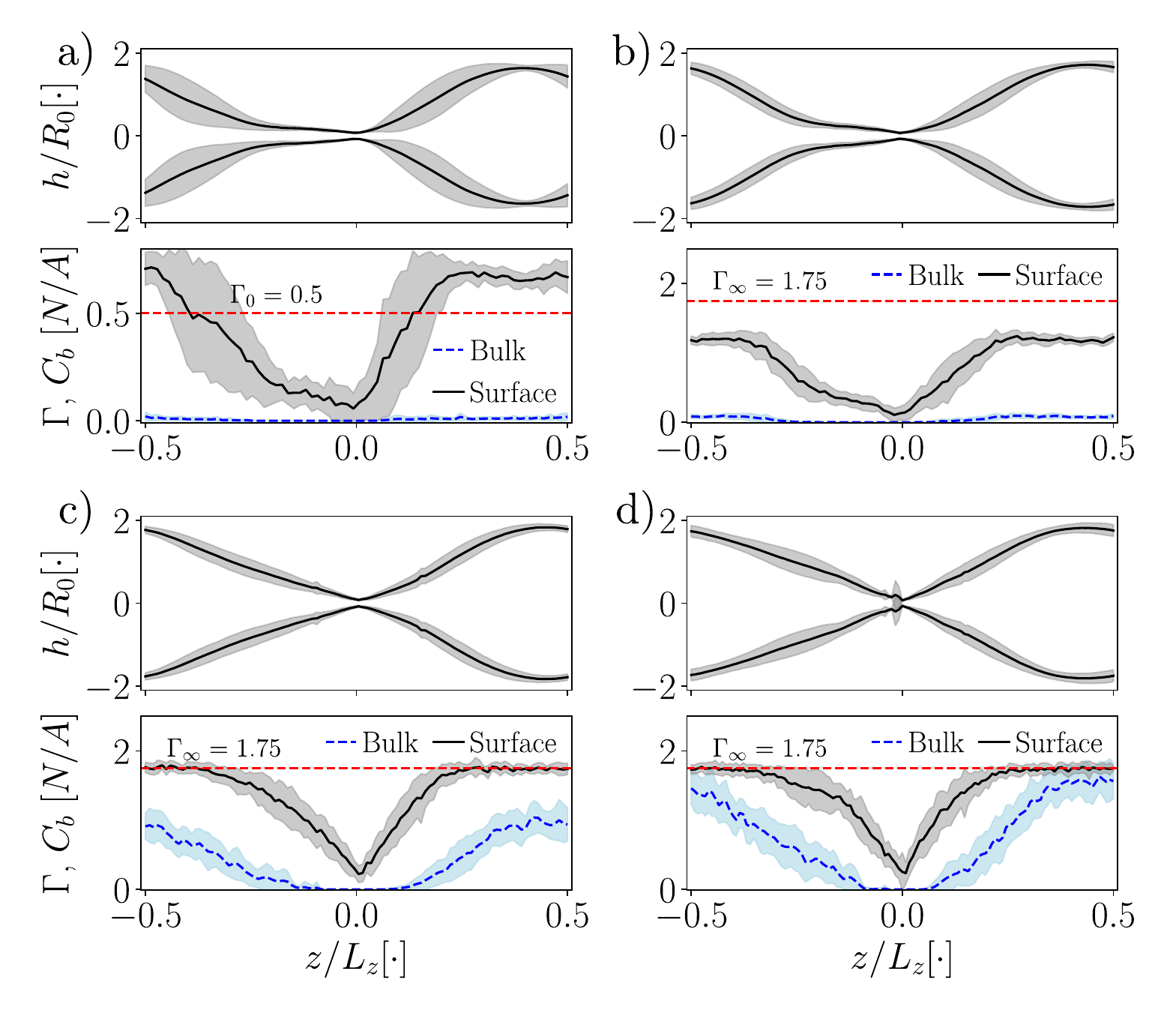}
\caption{\label{fig:shape-con} Profile shapes, $h/R_0$, and surfactant concentration along 
the $z$ axis at the surface and in the bulk of the fluid. Lines represent the ensemble
average and shaded area is one standard deviation. The averages were obtained 
at the moment of pinch-off and for two surfactant concentrations below CMC, a) $C=0.5$ and 
b) $C=1.0$, and two above CMC, c) $C=2.3$ and d) $C=2.9$. $\Gamma_{\infty}$ is the highest surfactant excess concentration at surface saturation. $\Gamma$
is the surface excess concentration and $C_b$ is the bulk concentration (number of 
molecules per surface area).
}
\end{figure}

Figure \ref{fig:shape-con} shows the breakup profile near the pinch-off of all 
the studied cases and the surfactant concentration distribution on the surface
and in the bulk phases. Because of the stochastic nature of the breakup, the averages
were taken by shifting the pinching point to $z=0$ and by flipping the profiles whenever
$\int_{-L/2}^0 hdz$ $>$ $\int^{L/2}_0 hdz$ to preserve large anti-symmetrical 
effects.\cite{zhao2020a} The lines show the ensemble-averaged profile and the 
shaded area is within one standard deviation.  In the instance of the lowest surfactant concentration, a prolonged neck emerges between droplets with an asymmetrical pinching point, deviating from the symmetrical double-cone profile predicted for thermal fluctuation breakup.
As surfactant concentration increases, an asymmetrical double cone profile begins to manifest, becoming almost symmetrical for the highest concentration. 

\begin{figure}[bt!]
\centering
\includegraphics[width=0.6\textwidth]{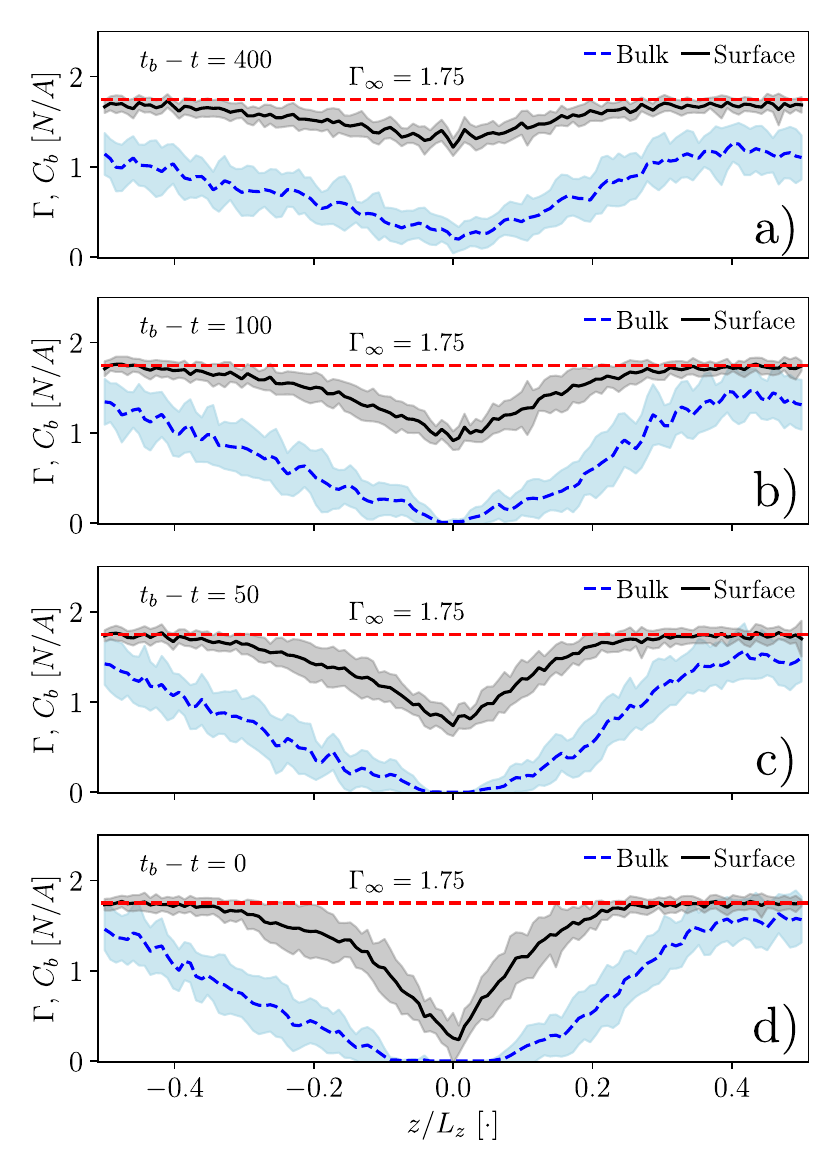}
\caption{\label{fig:con-time} Time evolution of the surfactant distribution for a case
above CMC ($C=2.9$) showing the advection of surfactants from the 
pinching point and increase in bulk surfactant in the main droplet region. $\Gamma$
is the surface excess concentration and $C_b$ is the bulk concentration (number of 
molecules per surface area). Each panel corresponds to a different
point in time before the breakup moment ($t=t_b$), i.e. 
a) $t_b-t=400$, b) $t_b-t=100$, c) $t_b-t=50$, and d) $t_b-t=0$,
as indicated. 
}
\end{figure}

The different panels of Fig. \ref{fig:shape-con} illustrate the distribution patterns of 
surfactant molecules within both the surface and bulk regions along the axis of the liquid 
thread. In the scenario where the initial concentration is at its lowest (panel a), 
an expected absence of surfactant molecules in the bulk phase is observed. 
This absence is attributed to the distance from the CMC. 
Furthermore, a discernible advection of surfactants is witnessed,
characterized by a reduction in surface concentration around the pinching 
point and a subsequent increase beyond its initial value of $\Gamma=0.5$ within the region 
occupied by the primary droplet. 
Contrary to this case, in all other instances, the presence of molecules within the bulk 
phase is evident. 
Both surface and bulk surfactants exhibit an advective motion away from the pinching point. 
In addition, it is noteworthy that the surface concentration consistently remains below
the saturation concentration $\Gamma_{\infty}$. 
An analysis of the temporal evolution, specifically examining the case with the highest initial 
concentration \blu{as} illustrated in Fig.~\ref{fig:con-time}, 
indicates a trajectory wherein the advected surface surfactants migrate toward
the main droplet region and subsequently infiltrate the bulk phase.
This observation suggests a directional flow pattern,
wherein surfactant molecules, once advected from the surface around the pinching point,
are transported towards the primary droplet region 
and eventually integrate into the bulk phase.

\subsection{Neck-radius scaling}

\begin{figure}[bt!]
\centering
\includegraphics[width=0.80\textwidth]{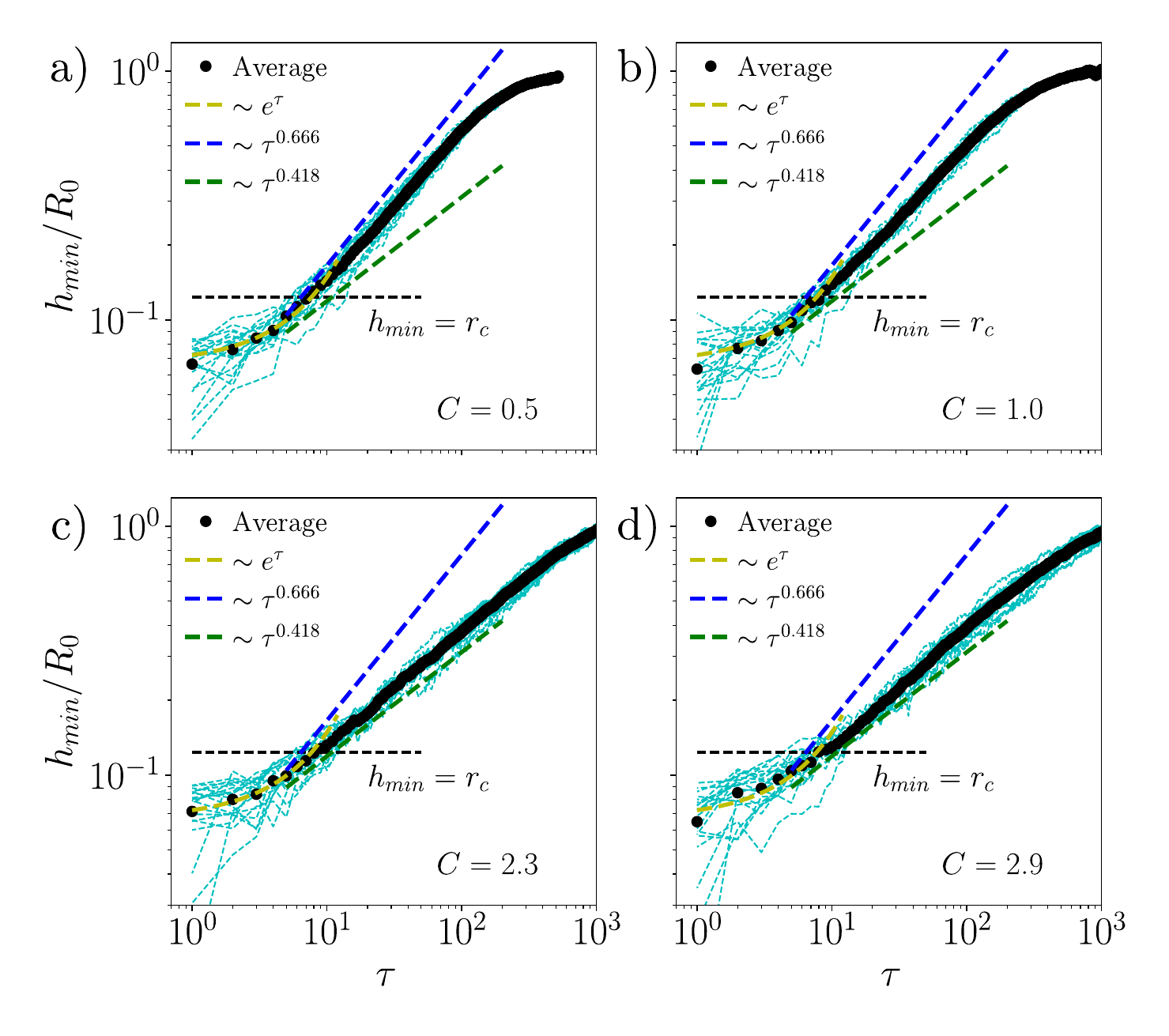}
\caption{\label{fig:scaling} Radius scaling of the thinning process for threads with four different
surfactant concentrations and comparison with known breakup regimes. a) $C = 0.5$; b) $C = 1.6$; c) $C = 2.3$; d) $C = 2.9$. Each case is an ensemble average of 20 simulations. $r_c$ is the cutoff distance for the attractive interaction. Dashed lines were added 
just as reference to the theoretical regimes and have the same values in all plots.
}
\end{figure}

Thinning dynamics of liquid threads under varying surfactant concentrations were 
systematically examined by tracking the minimum neck radius $h_{min}$ 
until the pinch-off event. 
Four distinct cases, encompassing two surfactant concentrations below the CMC and two above,
were investigated. 
For each case, 20 simulations were conducted, and the averaged
neck radius values were obtained with respect to their relative time from the breakup ($t_b$)
$\tau = (t_b - t)$ as presented in Fig.~\ref{fig:scaling}.
According to a theoretical analysis of surfactant-laden breakup 
at the limit when surfactants are advected away from the pinching point, 
the minimum radius should follow the viscous regime
with linear scaling $h_{min} \sim \tau$.\cite{wee2020}
However, we did not observe this scaling and at the lowest surfactant concentration, the thinning process 
appears to be close to
the inertial regime, indicated by the scaling $h_{min} \sim \tau^{2/3}$. 
Interestingly, for low concentration we did not observe a clear transition or the presence of the 
thermal fluctuation regime $h_{min} \sim \tau^{0.418}$.\cite{eggers2002}
MDPD simulations of a liquid without surfactants also have shown a deviation from the thermal
fluctuation scaling.\cite{zhao2021}

As we increase surfactant concentration, consequently reducing surface tension,
the breakup dynamics still exhibits a power-law behavior. However, the observed exponent
deviates from the inertial regime and approaches the thermal fluctuation one. 
These results substantiate the conclusions drawn by Zhao et al.,\cite{zhao2020a}
who demonstrated that the scaling exponent converges towards $0.418$ 
when the surface tension $\gamma \rightarrow 0$ from simulations using the 1D SLE model.
In all of our simulations cases, we can observe a transition between the power-law 
to a different regime immediately preceding the pinch-off event. 
This last regime has been proposed by Zhao et al. \cite{zhao2020}
as the ``breakup'' regime, following a power-law with exponent 0.
However, no further explanation was given as to the physical meaning and origin of this behavior.
Moreover, they only analyzed the evolution of a single simulation trajectory instead of an ensemble average. 
From our results, the transition actually occurs when $h_{min}$ reaches below the cutoff radius of 
the particle interactions and it could be an artifact from the MDPD method.
When the thread reaches such scales, the local density $\bar{\rho}$ (see Eq. \ref{eq4})
starts to decrease with time and this, in turn, reduces the strength of the repulsive term
in the conservative force, leading to a stronger relative attraction between the particles
that slows down the thinning process.
This possible regime scales not as power-law 
but seemingly as $h_{min} \sim e^{\tau}$. 
Exponential thinning is also found in the breakup of viscoelastic fluids\cite{chang1999} 
and when surface rheological effects are taken into account.\cite{martinez-calvo2020,wee2022}

\section{Conclusions}
\label{conclusions}

In this study, we investigated the impact of surfactants on the breakup of a liquid 
thread under thermal fluctuations by employing simulations via the particle-based 
mesoscale method MDPD.
We characterized the HT3 coarse-grained surfactant model in terms of its relevant
interfacial properties such as surface tension and interface thickness.
Furthermore, the model also features the spontaneous formation of micelles above the CMC.
The results indicate that the model is adequate in reproducing the behavior 
of real surfactant molecules. Then, a detailed study of the breakup of liquid threads
in the presence of surfactant was carried out.
We find that increased surfactant concentration would naturally lead to the
increase in the most unstable wavelength that characterizes the breakup, 
and, also, in the time for it to occur. 
At first, this could be explained by the decrease in surface tension which, in turn, 
increases the Oh number. However, both the wavelength and time to breakup 
keep increasing for concentrations above the CMC, where surface tension remains 
constant. This might therefore be due to bulk viscous effects from the increase in 
surfactant micelle concentration.
We also discussed the minimum radius scaling of the thinning process and 
showed that the observed
power-law behavior matches the I regime and approaches the TF regime, depending on the surfactant concentration --- low and high, respectively.
At low concentrations, we found that  
the breakup is in the I regime instead of the universal VI or V regimes. 
Nonetheless, this result is in agreement with experimental measurements.\cite{kovalchuk2016} 
At higher concentrations, the thinning dynamics transition
into the thermal fluctuation regime.
Moreover, we argue that the ``breakup'' thinning regime proposed 
by Zhao et al. \cite{zhao2020a}
might be an artifact of the MDPD model as the transition into it occurs when 
$h_{min}$ approaches the cutoff of the interactions. When the minimum radius is
below the cutoff, the attraction between particles becomes increasingly more 
dominant than the repulsion leading to the slow-down of the thinning dynamics.
Finally, we have examined the breakup profiles and surfactant distribution of the
threads providing molecular-level insights into the mechanisms of this phenomenon.
Thus, we anticipate that our study sheds light on the thus far poorly investigated
breakup of liquid threads laden with surfactant.

\begin{acknowledgments}
This research has been supported by the National 
Science Centre, Poland, under
grant No.\ 2019/34/E/ST3/00232. 
We gratefully acknowledge Polish high-performance 
computing infrastructure PLGrid (HPC Centers: ACK Cyfronet AGH) 
for providing computer facilities and support 
within computational grant no. PLG/2022/015747.
\end{acknowledgments}

\bibliography{aipsamp}

\end{document}